\documentclass[a4paper]{spie}  
\pdfoutput=1

\usepackage{array,booktabs,multicol,pgfplotstable}
\usepackage{algorithm2e}
\SetAlCapSty{}
\usepackage{tikz,pgfplots}
\pgfplotsset{compat=1.13}
\usepgfplotslibrary{fillbetween}
\usepgfplotslibrary{polar}
\usepackage{wrapfig}
\usepackage[]{todonotes}
\usepackage{enumitem}
\usepackage{subcaption}
\usepackage{setspace}

\usepackage{amsmath,amsfonts,amssymb}
\usepackage{mathtools}
\usepackage{graphicx}
\usepackage{cleveref}

\usepackage[style=numeric-comp]{biblatex}
\AtEveryBibitem{
  \clearfield{month}
  \clearfield{series}
  \clearfield{venue}
  \clearname{editor}
  \clearlist{publisher}
  \clearlist{organization}
  \clearlist{location} 
  \clearfield{doi}
  \clearfield{url}
  \clearfield{issn}
  \clearfield{isbn}
  \clearfield{urldate}
  \clearfield{eventdate}
  \clearfield{note}
}
\DefineBibliographyStrings{english}{%
  references = {REFERENCES},
}
\newcommand{\cf}{\textit{c.f.}}
\newcommand{\eg}{\textit{e.g.}}

\newcommand{\ie}{\textit{i.e.}}

\newcommand{\invivo}{\textit{in vivo}}

\DeclareMathOperator*{\argmin}{arg\,min}

\title{Imaging the Developing Heart: Synchronized Timelapse Microscopy During Developmental Changes}

\author[a]{Carl J. Nelson}
\author[b]{Charlotte Buckley}
\author[b]{John J. Mullins}
\author[b]{Martin A. Denvir}
\author[a]{Jonathan Taylor}

\affil[a]{School of Physics and Astronomy, University of Glasgow, Glasgow, UK}
\affil[b]{Centre for Cardiovascular Science, Queens Medical Research Institute, University of Edinburgh, Edinburgh, UK}

\authorinfo{Further author information: (Send correspondence to J.T. and C.J.N.)\\J.T.: E-mail: jonathan.taylor@glasgow.ac.uk;~~C.J.N.: E-mail: chas.nelson@glasgow.ac.uk}

\begin{document}
\maketitle

\begin{abstract}
How do you use imaging to analyse the development of the heart, which not only changes shape but also undergoes constant, high-speed, quasi-periodic changes? We have integrated ideas from prospective and retrospective optical gating to capture long-term, phase-locked developmental time-lapse videos. In this paper we demonstrate the success of this approach over a key developmental time period: heart looping, where large changes in heart shape prevent previous prospective gating approaches from capturing phase-locked videos. We use the comparison with other approaches to \invivo{} heart imaging to highlight the importance of collecting the most appropriate data for the biological question.
\end{abstract}

\keywords{zebrafish, heart, development, optical gating, light sheet, SPIM}


\section{INTRODUCTION}\label{sec:intro}


Understanding the molecular basis of heart development is essential to understanding and developing treatments for congenital heart diseases. Thanks to external development, transparent embryos and powerful transgenics the zebrafish has become an invaluable vertebrate model for the study of heart development, as reviewed in~\cite{BSQetal2016}. Though simpler than the human heart (with only two chambers), the zebrafish has many similarities and many models of congenital heart diseases are now available.

A key stage in heart development is heart looping --- the morphological transition from a linear heart tube to a multi-chambered heart. Heart looping is a dynamic process and 3D imaging of the heart allows quantitative analyses of heart development during this crucial stage. However, building up 3D videos of the heart usually requires the sequential capture of 2D images. In living zebrafish this is impeded by the constant motion of the beating heart and, as such, constructing 3D videos that are `frozen' at one point in the heartbeat is necessary.

%

\subsection{Retrospective Optical Gating}

One solution to this problem is to computationally post-process a video spanning at least one heartbeat, assigning frames to known heartbeat phases --- a technique termed retrospective gating, \eg{}\cite{LFGetal2005,LVFetal2006}. In order to reconstruct 3D data of the heart, videos spanning at least one heartbeat are captured at multiple planes. These videos are then temporally aligned to `phase match' the different slices. From this a 3D reconstruction can be extracted.

As the acquired raw videos contain information about entire heart periods, retrospective optical gating approaches are particularly suited to capturing 3D dynamics throughout a heartbeat, such as calcium dynamics~\supercite{MSWetal2014}. However, if the requirement is a heart `frozen' at one phase, \ie{} for long-term developmental imaging, significantly more fluorescence images are captured than will be used leading to damaging levels of bleaching or photodamage/toxicity in the sample.

\subsection{Prospective Optical Gating}

Alternatively, our approach of prospective optical gating aims to predictively trigger image acquisition, in a manner similar to using a real-time ECG, to capture frames only at the desired phase in the heart cycle. We have previously shown how a brightfield channel can be used to track heart cycle phase and predictively trigger fluorescence image capture on a light sheet microscope to provide high quality, synchronised 3D stacks\supercite{TGL2012,TSLetal2011}.

This approach relies on a set of brightfield reference frames for `phase matching'. During development the heart undergoes morphological changes and brightfield frames early in development do not bear close resemblance to frames from later in development. This is particularly true for heart looping where the heart undergoes a massive morphological change. New approaches must be developed that can update these reference frames and maintain `phase locked' 3D image capture over developmental time scales.

In this paper we describe a new approach for updating reference frames over relatively long time periods that allows for constant phase locked 3D image capture over developmental time scales. This approach builds upon retrospective gating concepts to provide a robust system for imaging the beating zebrafish heart through development and also through repair and regeneration.

\section{METHODS \& RESULTS}\label{sec:methods}


\subsection{Real-Time Phase Matching by Prospective Gating}

For real-time phase matching we use prospective gating as described in~\cite{TGL2012}. Briefly, a high frame rate (80 fps) brightfield channel is compared to a reference complete heartbeat and assigned a phase. Forward prediction is then carried out to accurately trigger the acquisition of a fluorescence image, one per heartbeat, at the desired target phase. Fluorescent images are captured at the desired phase for each plane within the heart to build up 3D data of the computationally frozen heart.

\begin{figure}
  \begin{subfigure}{0.4\linewidth}
    \centering
    \includegraphics[width=0.48\linewidth]{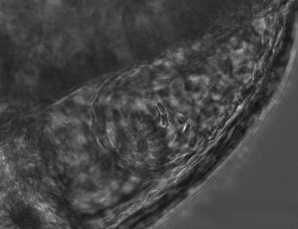}
    \hfill
    \includegraphics[width=0.48\linewidth]{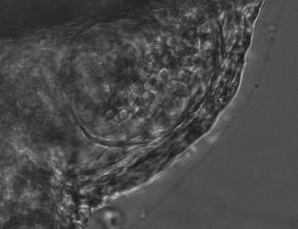}
    \caption{}\label{fig:scc:bf}
  \end{subfigure}
  \begin{subfigure}{0.6\linewidth}
    \centering
    \begin{tikzpicture}%
        \pgfplotsset{set layers}
        \begin{axis}[
          scale only axis,
          width=0.8\linewidth,
          height=0.4\linewidth,
          grid=major, 
          grid style={dashed,gray!30}, 
          xlabel= Time/minutes, 
          ylabel= Error in Target Frame/Radians,
          legend style={at={(0.5,1.05)},anchor=south,font=\tiny},
          legend columns = 4,
          xmin=0, xmax=111.3,
          ymin=-3.15, ymax=+3.15,
          ytick = {-3.14159,-1.57080,0,1.57080,3.14159},
          yticklabels = {$-\pi$,$-\frac{\pi}{2}$,0,$-\frac{\pi}{2}$,$-\pi$},
          enlargelimits=false,
          ]
          \addplot[name path=pluserror, draw=none, no markers, forget plot] coordinates {
            (0.0,0.08675912279322029)
            (8.560861510704793,0.08475470146372605)
            (17.121723021409526,0.08757142187544575)
            (25.68258453211432,0.08951298457979823)
            (34.24344604281905,0.0900441947716894)
            (42.80430755352384,0.09027956271083881)
            (51.365169064228574,0.09090846451859064)
            (59.92603057493337,0.08644375340566215)
            (68.4868920856381,0.0858480532985689)
            (77.0477535963429,0.08671627980024023)
            (85.60861510704763,0.08732994744775119)
            (94.16947661775242,0.0897180607719739)
            (102.73033812845715,0.0850093249021262)
            (111.29119963916195,0.08407512024458581)
          };
          \addplot[name path=minuserror, draw=none, no markers, forget plot] coordinates {
            (0.0,-0.08675912279322029)
            (8.560861510704793,-0.08475470146372605)
            (17.121723021409526,-0.08757142187544575)
            (25.68258453211432,-0.08951298457979823)
            (34.24344604281905,-0.0900441947716894)
            (42.80430755352384,-0.09027956271083881)
            (51.365169064228574,-0.09090846451859064)
            (59.92603057493337,-0.08644375340566215)
            (68.4868920856381,-0.0858480532985689)
            (77.0477535963429,-0.08671627980024023)
            (85.60861510704763,-0.08732994744775119)
            (94.16947661775242,-0.0897180607719739)
            (102.73033812845715,-0.0850093249021262)
            (111.29119963916195,-0.08407512024458581)
          };
          \addplot[fill=black, opacity=0.5, forget plot] fill between[on layer={}, of=pluserror and minuserror];
          \addplot[color=black,draw=none, no markers] coordinates {
            (0.0,0)
            (8.560861510704793,0)
            (17.121723021409526,0)
            (25.68258453211432,0)
            (34.24344604281905,0)
            (42.80430755352384,0)
            (51.365169064228574,0)
            (59.92603057493337,0)
            (68.4868920856381,0)
            (77.0477535963429,0)
            (85.60861510704763,0)
            (94.16947661775242,0)
            (102.73033812845715,0)
            (111.29119963916195,0)
          };
          \addlegendentry{Manually Updated}
          \addplot[only marks,color=red] coordinates {
            (0.0,0.0)
            (8.560861510704793,-0.0016185793682303995)
            (17.121723021409526,0.020353025175168682)
            (25.68258453211432,0.11458267236875663)
            (34.24344604281905,0.1782138524308854)
            (42.80430755352384,0.1756713006480881)
            (51.365169064228574,-0.10137020073374092)
            (59.92603057493337,2.9446467787490316)
            (68.4868920856381,-0.6914922537371737)
            (77.0477535963429,-0.9607674060923967)
            (85.60861510704763,-0.8773129892541753)
            (94.16947661775242,-0.46184988932118465)
            (102.73033812845715,-0.01608242384960956)
            (111.29119963916195,-0.011303355504447055)
          };
          \addlegendentry{No Update}
          \addplot[only marks,color=blue] coordinates {
            (0.0,0.0)
            (8.560861510704793,-0.0858464127950187)
            (17.121723021409526,-0.037576463008947236)
            (25.68258453211432,0.11762266128284082)
            (34.24344604281905,-0.0028610467347567337)
            (42.80430755352384,-0.004279897837881741)
            (51.365169064228574,0.1253343565464604)
            (59.92603057493337,0.4404604691120162)
            (68.4868920856381,-0.11064214566402875)
            (77.0477535963429,-0.15087377233357913)
            (85.60861510704763,-0.020525605986491335)
            (94.16947661775242,0.10416081694868096)
            (102.73033812845715,-0.21794492154929124)
            (111.29119963916195,-0.22108396571320643)
          };
          \addlegendentry{Updated}
        \end{axis}
    \end{tikzpicture}
    \caption{}\label{fig:scc:scc}
  \end{subfigure}
  \caption{\subref{fig:scc:bf} Brightfield images of the developing zebrafish heart at 2.5 dpf (left) and the same heart, at the same heartbeat phase, a day later at 3.5 dpf (right). \subref{fig:scc:scc} Without updating the reference heartbeat, changes in heart shape, size and beat lead to errors in maintaining the synchronisation at a specific target phase (red markers); however, using retrospective gating to update the reference heartbeat minimises these errors (blue markers). The grey band indicates the manually determined target frame for each update ($\pm1$ frame). All values are relative to the manually determined target frame and converted into radians for visualisation.}\label{fig:scc}
\end{figure}
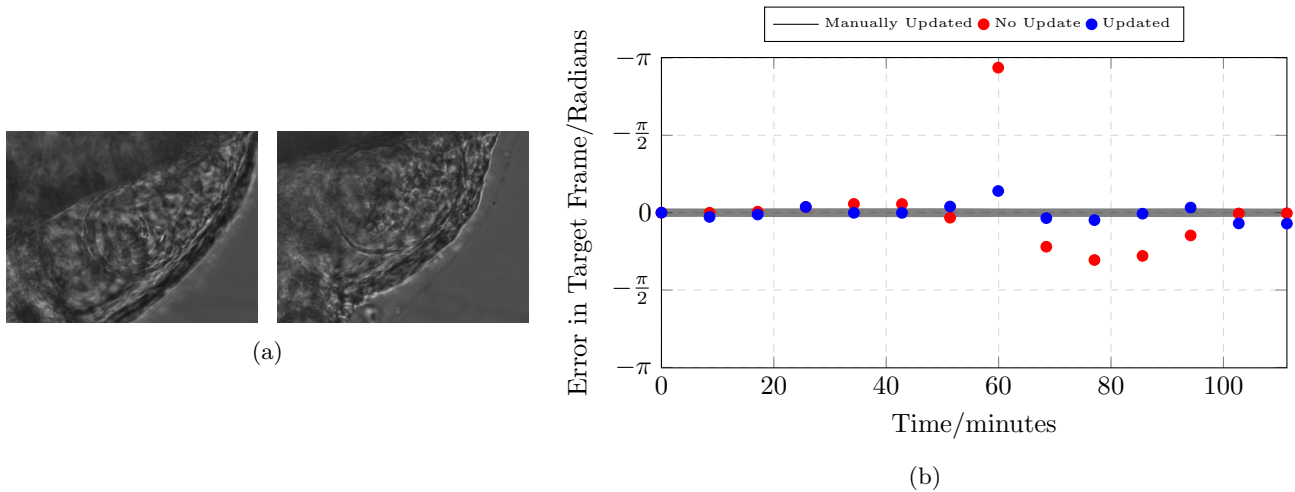

However, throughout development the zebrafish heart changes in size, shape and beat rate, causing difference between live brightfield images and the original reference heartbeat. These differences lead to errors in the prospective optical gating that will get ever larger when capturing over long, developmental time periods (\cref{fig:scc}).

\subsection{Long-Term Phase Locking by Retrospective Gating}

In order to successfully acquire fluorescence images at a constant target phase over long periods of time, we need to be able to update the reference heartbeat images at regular intervals. Updating the reference heartbeat allows the system to cope with changes in heart size and morphology that occur throughout development. However, whilst selecting a new reference heartbeat can be done as described in~\cite{TSLetal2011}, we require a way to determine the equivalent frame in the new reference heartbeat that matches the original, user-defined, target heart phase.

In order to phase lock across updated reference heartbeats, we use inter-frame correlation\supercite{LFGetal2005} between the current reference heartbeat, $R_i$, and a proposed new reference heartbeat, $R_j$. First, reference heartbeats are re-sampled so that all reference hearts beats have an identical, integer number of frames ($R^N_i$ and $R^N_j$). Then the relative phase shift between reference heartbeats, $\Delta\phi$, is determined as the minimum value cross correlation between both sequences (computed in Fourier space for speed),
\begin{equation}
  \Delta\phi_{i,j} = \argmin_{\phi \in N}{\left(R^N_i \star R^N_j\right)}.\label{eq:Deltaphi}
\end{equation}
To refine this shift to sub-frame precision we use V-fitting as described in~\cite{TSLetal2011}. This temporal alignment allows for identification of the relative phase shift and hence the new target frame in the new reference heartbeat. Note that this phase shift is in the range $[0,N)$ and $\Delta\phi_{i,i} = 0$.

In principle, $\Delta\phi$ between all adjacent sequences can be used to determine the global phase shift as follows,
\begin{equation}
  \Phi_k = \sum_{i=0}^{k-1}{\Delta\phi_{i,i+1}}.\label{eq:Phik}
\end{equation}
This global phase shift determines the position in $R_k$ that matches the target frame for synchronization in $R_0$.

\subsection{Drift Correction}

\begin{figure}
  \centering
  \begin{tikzpicture}%
      \pgfplotsset{set layers}
      \begin{axis}[
        scale only axis,
        width=0.8\linewidth,
        height=0.24\linewidth,
        grid=major, 
        grid style={dashed,gray!30}, 
        xlabel= Time/minutes, 
        ylabel= Error in Target Frame/Radians,
        legend style={at={(0.5,1.05)},anchor=south,font=\tiny},
        legend columns = 4,
        xmin=0, xmax=12.02,
        ymin=-3.15, ymax=+3.15,
        axis y line*=left,
        enlargelimits=false,
        ]
        \addplot[name path=pluserror, draw=none, no markers, forget plot] coordinates {
          (0.0,0.1899596488136419)
          (1.0004199594666716,0.18626747991911377)
          (2.0020899230476212,0.18896797916329583)
          (3.003343218590483,0.18720539879055056)
          (4.005221516190477,0.18102164072581764)
          (5.006266477714295,0.1871273518070541)
          (6.007728107276201,0.1854209968862651)
          (7.008981402819048,0.19507128760782325)
          (8.010443032380953,0.18956206551609825)
          (9.011904661942859,0.18868465505108614)
          (10.013366291504765,0.20175805511591985)
          (11.014619587047628,0.19136940502274968)
          (12.015872882590475,0.19143762075567458)
        };
        \addplot[name path=minuserror, draw=none, no markers, forget plot] coordinates {
          (0.0,-0.1899596488136419)
          (1.0004199594666716,-0.18626747991911377)
          (2.0020899230476212,-0.18896797916329583)
          (3.003343218590483,-0.18720539879055056)
          (4.005221516190477,-0.18102164072581764)
          (5.006266477714295,-0.1871273518070541)
          (6.007728107276201,-0.1854209968862651)
          (7.008981402819048,-0.19507128760782325)
          (8.010443032380953,-0.18956206551609825)
          (9.011904661942859,-0.18868465505108614)
          (10.013366291504765,-0.20175805511591985)
          (11.014619587047628,-0.19136940502274968)
          (12.015872882590475,-0.19143762075567458)
        };
        \addplot[fill=black, opacity=0.5, forget plot] fill between[on layer={}, of=pluserror and minuserror];
        \addplot[color=black,draw=none, no markers] coordinates {
          (0.0,0)
          (1.0004199594666716,0)
          (2.0020899230476212,0)
          (3.003343218590483,0)
          (4.005221516190477,0)
          (5.006266477714295,0)
          (6.007728107276201,0)
          (7.008981402819048,0)
          (8.010443032380953,0)
          (9.011904661942859,0)
          (10.013366291504765,0)
          (11.014619587047628,0)
          (12.015872882590475,0)
        };
        \addlegendentry{Manually Updated}
        \addplot[only marks,color=red] coordinates {
          (0.0,0.0)
          (1.0004199594666716,-1.3797519909722817)
          (2.0020899230476212,-2.229084574168088)
          (3.003343218590483,-3.064233348690947)
          (4.005221516190477,1.802055561864985)
          (5.006266477714295,1.6740678317266093)
          (6.007728107276201,0.9498439655299749)
          (7.008981402819048,0.9217318900146942)
          (8.010443032380953,0.8011268805239617)
          (9.011904661942859,0.7454077658503113)
          (10.013366291504765,0.7287082352802825)
          (11.014619587047628,0.6656363400956637)
          (12.015872882590475,0.685076328017522)
        };
        \addlegendentry{No Drift Correction}
        \addplot[only marks,color=blue] coordinates {
          (0.0,0.0)
          (1.0004199594666716,-0.12556849875434484)
          (2.0020899230476212,0.2436958716049813)
          (3.003343218590483,0.508257518254509)
          (4.005221516190477,-0.0015252407413061941)
          (5.006266477714295,-0.10349824606590508)
          (6.007728107276201,-0.0191700505611907)
          (7.008981402819048,-0.011265056857537203)
          (8.010443032380953,-0.05723285693760993)
          (9.011904661942859,-0.03861600119118602)
          (10.013366291504765,0.3569580311436402)
          (11.014619587047628,0.31398019422335377)
          (12.015872882590475,0.34461918684963866)
        };
        \addlegendentry{With Drift Correction}
        \addplot[color=green,draw=none] coordinates {(0,0)};
        \addlegendentry{Magnitude of Drift}
      \end{axis}
      \begin{axis}[
        scale only axis,
        width=0.8\linewidth,
        height=0.24\linewidth,
        ylabel= Magnitude of Drift/pixels,
        xmin=0, xmax=12.02,
        ymin=0, ymax=9,
        axis y line*=right,
        enlargelimits=false,
        ]
        \addplot[no markers,color=green,semithick] coordinates {
          (2.0020899230476212,8.06225774829855)
          (3.003343218590483,7.0)
          (4.005221516190477,8.0)
          (5.006266477714295,6.0)
          (6.007728107276201,6.0)
          (7.008981402819048,5.0)
          (8.010443032380953,6.0)
          (9.011904661942859,4.0)
          (10.013366291504765,4.0)
          (11.014619587047628,4.0)
          (12.015872882590475,1.4142135623730951)
        };
      \end{axis}
  \end{tikzpicture}
  \caption{Drift correction is important for the accurate determination of relative phase shift, $\Delta\phi_{i,j}$. Without drift correction (red markers) errors accumulate over time; however, with drift correction (blue markers) drift-based errors are minimised. The grey band indicates the manually determined target frame for each update ($\pm1$ frame). All values are relative to the manually determined target frame and converted into radians for visualisation.}\label{fig:drift}
\end{figure}
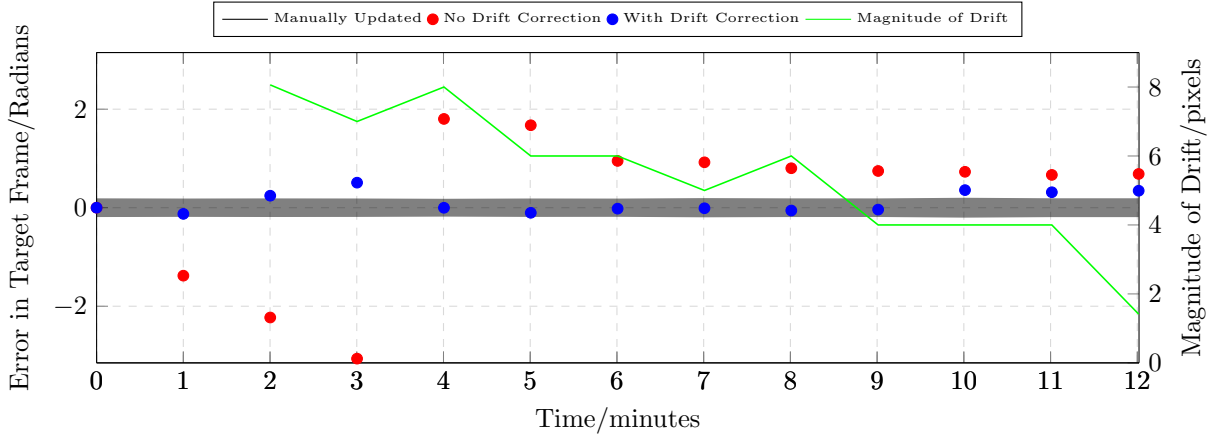

One major source of error in the calculation of $\Delta\phi$ is motion of the sample in the $xy$ plane. This may be due to growth or movement of the fish within the sample holder. Without correction for this drift, the cross correlation will lose effectiveness due to a reduction in pixelwise similarity. This will lead to errors in $\Delta\phi_{i,j}$ which accumulate over time (\cref{fig:drift}). Using the drift tracking incorporated into the real-time phase matching, as introduced in~\cite{TGL2012}, we are able to correct for drift between pairs of reference heartbeats. With this correction such drift-based errors are minimised as demonstrated in \cref{fig:drift}.

\subsection{Minimising Accumulated Errors with Least-Squares Regression}

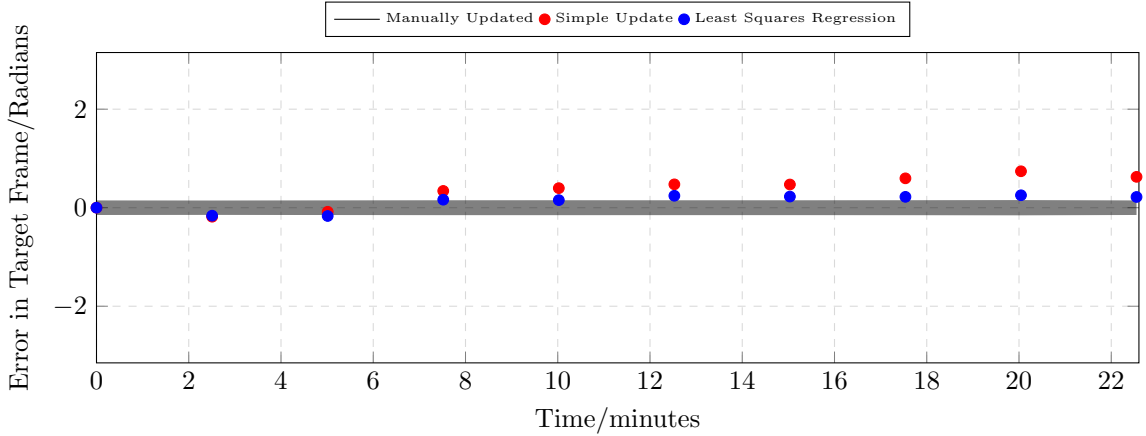
\begin{figure}
  \centering
  \begin{tikzpicture}%
      \pgfplotsset{set layers}
      \begin{axis}[
        scale only axis,
        width=0.8\linewidth,
        height=0.24\linewidth,
        grid=major, 
        grid style={dashed,gray!30}, 
        xlabel= Time/minutes, 
        ylabel= Error in Target Frame/Radians,
        legend style={at={(0.5,1.05)},anchor=south,font=\tiny},
        legend columns = 4,
        xmin=0, xmax=22.6,
        ymin=-3.15, ymax=+3.15,
        enlargelimits=false,
        ]
        \addplot[name path=pluserror, draw=none, no markers, forget plot] coordinates {
          (0.0,0.1479019316343306)
          (2.5068832512000045,0.14611666210959504)
          (5.012308164266665,0.14886178345765003)
          (7.517648848152385,0.15122714410212518)
          (10.023073761219045,0.15173101610536557)
          (12.528082006247617,0.15097471337446305)
          (15.032701411657142,0.1508145879071392)
          (17.537709656685713,0.1520896537350056)
          (20.042926235733336,0.15488520936539654)
          (22.547943411200002,0.1478133577148584)
        };
        \addplot[name path=minuserror, draw=none, no markers, forget plot] coordinates {
          (0.0,-0.1479019316343306)
          (2.5068832512000045,-0.14611666210959504)
          (5.012308164266665,-0.14886178345765003)
          (7.517648848152385,-0.15122714410212518)
          (10.023073761219045,-0.15173101610536557)
          (12.528082006247617,-0.15097471337446305)
          (15.032701411657142,-0.1508145879071392)
          (17.537709656685713,-0.1520896537350056)
          (20.042926235733336,-0.15488520936539654)
          (22.547943411200002,-0.1478133577148584)
        };
        \addplot[fill=black, opacity=0.5, forget plot] fill between[on layer={}, of=pluserror and minuserror];
        \addplot[color=black,draw=none, no markers] coordinates {
          (0.0,0)
          (2.5068832512000045,0)
          (5.012308164266665,0)
          (7.517648848152385,0)
          (10.023073761219045,0)
          (12.528082006247617,0)
          (15.032701411657142,0)
          (17.537709656685713,0)
          (20.042926235733336,0)
          (22.547943411200002,0)
        };
        \addlegendentry{Manually Updated}
        \addplot[only marks,color=red] coordinates {
          (0.0,0.0)
          (2.5068832512000045,-0.18302020931309126)
          (5.012308164266665,-0.0824206295186034)
          (7.517648848152385,0.34067116208335424)
          (10.023073761219045,0.3967424175995067)
          (12.528082006247617,0.47409661140331905)
          (15.032701411657142,0.4710101893313441)
          (17.537709656685713,0.5961957446263231)
          (20.042926235733336,0.7395925274086488)
          (22.547943411200002,0.6264151602128645)
        };
        \addlegendentry{Simple Update}
        \addplot[only marks,color=blue] coordinates {
          (0.0,4.440892098500626e-16)
          (2.5068832512000045,-0.1620374106485145)
          (5.012308164266665,-0.16819414073306493)
          (7.517648848152385,0.16236684349753916)
          (10.023073761219045,0.1539930704252921)
          (12.528082006247617,0.24238089303761468)
          (15.032701411657142,0.22593721530298128)
          (17.537709656685713,0.22041714747794794)
          (20.042926235733336,0.2526270458841775)
          (22.547943411200002,0.21804170984867244)
        };
        \addlegendentry{Least Squares Regression}
      \end{axis}
  \end{tikzpicture}
  \caption{Least squares regression minimises accumulated errors in $\Phi_k$. Without least squares regression (red markers) errors accumulate with time; however, with least squares regression ($\Phi'_k$; blue markers) these errors do not accumulate. The grey band indicates the manually determined target frame for each update ($\pm1$ frame). All values are relative to the manually determined target frame and converted into radians for visualisation.}\label{fig:mcc}
\end{figure}

Unfortunately, in practice, even after drift correction, accumulated, small, random errors mean that~\cref{eq:Phik} is not robust enough for phase locking over developmental time scales, \eg{} over a 24 hour period. Across multiple synchronization resets $\Phi_k$ will include the accumulated errors of all $\Delta\phi$ and the synchronization will drift away from the target heart phase (\cref{fig:mcc}).

We protect against this accumulated error by incorporating relative phase shifts between historical reference heartbeats into a weighted, linear least-squares regression to determine a corrected global phase shift. This is analogous to the approach used in~\cite{LFGetal2005} for aligning planes. Reference heartbeats separated further in time are given a lower weighting than those temporally close together as determined by their cross correlation values,
\begin{equation}
  w_{i,j}=\min{\left(R^N_i \star R^N_j\right)}.
\end{equation}
From the resultant, overdetermined system of equations, we solve for the corrected global phase shift, $\Phi'_k$, that most closely satisfies \textit{all} the relative shift comparisons, $\Delta\phi_{i,j}$ that we have made (\cref{fig:mcc}).

Solving this system of equations is actually non-trivial, since each equation for $\Delta\phi_{i,j}$ has a modulo $N$ solution and standard linear algebra solvers are not compatible with modular arithmetic. If the equations were to be solved naively, \ie{} without modular arithmetic, the equations might be found to be self-contradictory.

In order to address this issue, we precondition our equations by first solving only for nearest-neighbours, $\Delta\phi_{i,i+1}$, \cf{}~\cref{eq:Phik}. This gives an estimated solution but one that is subject to the accumulated errors described above. However, this estimate allows us to `unwrap' all relative phase shifts, $\Delta\phi_{i,j}$, into a self-consistent, global and non-modulo frame of reference, $\Phi'_k \in [0,\infty)$ where $\Phi'_0 = \Phi_0 = 0$. These global phase shifts are then used to yield a correct target frame index in $R'_k$ that is less susceptible to random variation over multiple synchronization resets.

\begin{figure}
  \centering
  \begin{subfigure}{0.3\linewidth}
    \includegraphics[width=\linewidth]{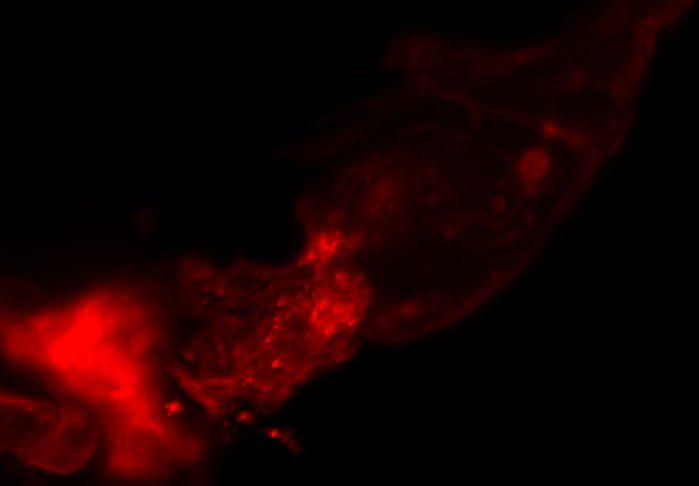}
    \caption{0 hours}
  \end{subfigure}
  \begin{subfigure}{0.3\linewidth}
    \includegraphics[width=\linewidth]{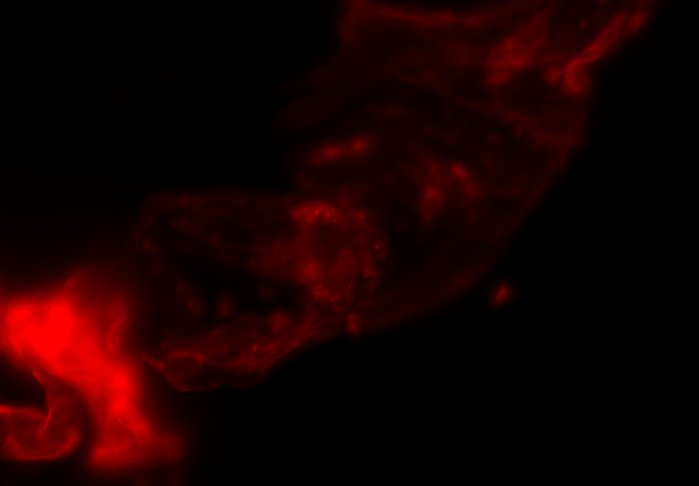}
    \caption{4 hours}
  \end{subfigure}
  \begin{subfigure}{0.3\linewidth}
    \includegraphics[width=\linewidth]{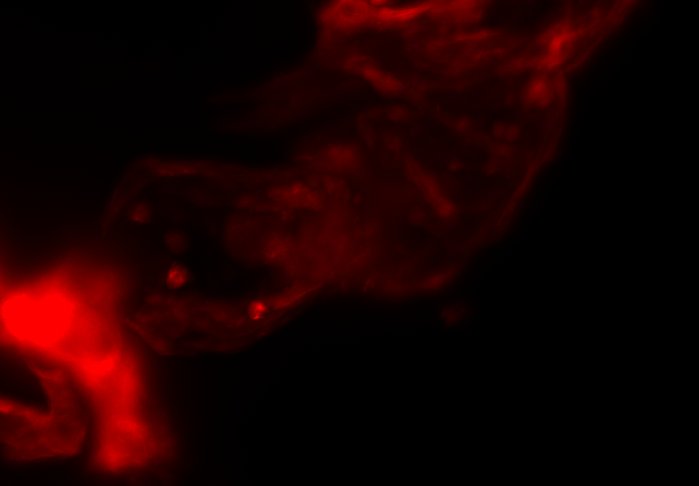}
    \caption{8 hours}
  \end{subfigure}

  \begin{subfigure}{0.3\linewidth}
    \includegraphics[width=\linewidth]{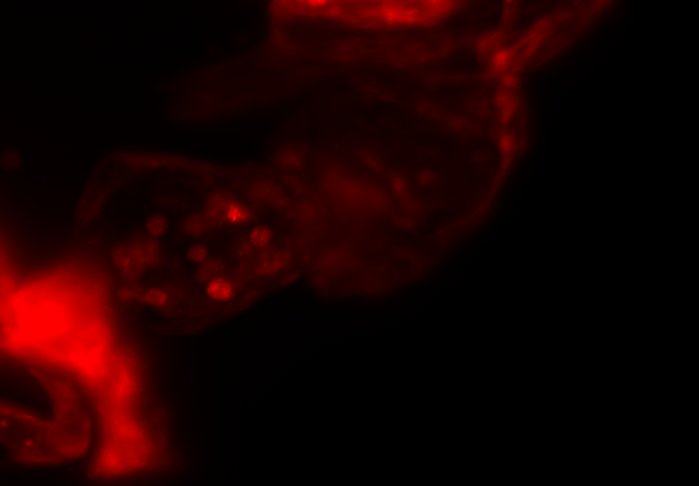}
    \caption{12 hours}
  \end{subfigure}
  \begin{subfigure}{0.3\linewidth}
    \includegraphics[width=\linewidth]{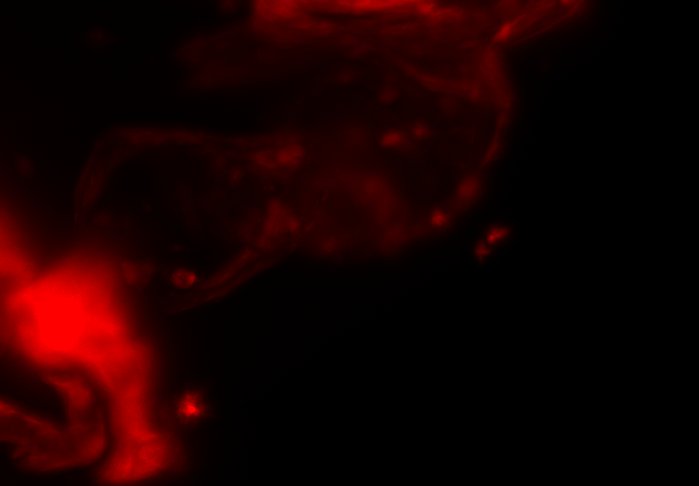}
    \caption{16 hours}
  \end{subfigure}
  \begin{subfigure}{0.3\linewidth}
    \includegraphics[width=\linewidth]{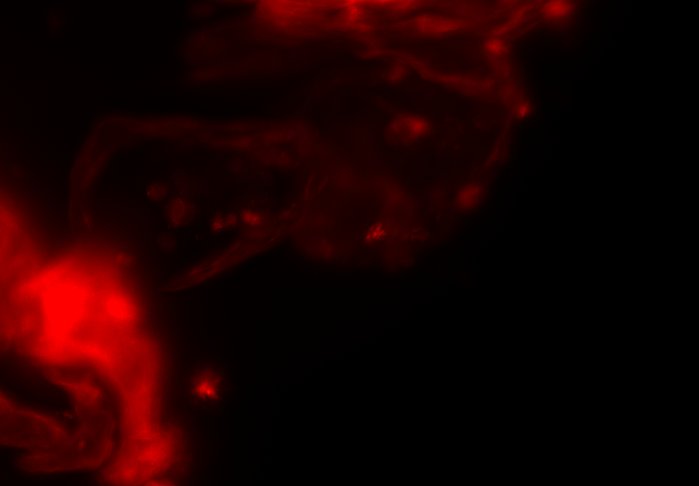}
    \caption{18 hours}
  \end{subfigure}
  \caption{Combining propsective and retrospective optical gating techniques allows for phase-locked imaging over developmental time scales. (a-f) Maximum intensity projections of 3D phase-matched stacks of the developing zebrafish heart (2.5 dpf - 3.5 dpf).}\label{fig:fluor}
\end{figure}

The combination of prospective optical gating for real-time phase matching and prospective optical gating for long-term phase locking allow us to capture in phase images over extended periods of times. \Cref{fig:fluor} demonstrates this phase locking over 18+ hours across a key developmental phase - heart looping.

\section{CONCLUSION}\label{sec:discuss}


We have previously shown the use of prospective optical gating with SPIM for the capture of 3D fluorescence data of the living, beating zebrafish heart. However, until now it was not possible to maintain synchronisation over timescales of an hour or more, due to changes in the appearance of the heart. Here we have demonstrated the successful application of combining prospective and retrospective optical gating techniques in order to update the reference heartbeat and phase lock between old and new references.

We have described how a combination of drift correction, inter-frame correlation and least squares regression can be used to accurately and robustly update the reference heartbeat and target frame. We have shown that this allows for phase locked imaging over developmental time scales. In particular we have demonstrated the ability to maintain synchronisation over 18+ hours around the heart looping stage of cardiac development in zebrafish.

This approach for updating reference heartbeats allow our real-time prospective gating system to be robust to changes in heart size, shape and heart rate. However, using inter-frame correlation does not allow for changes in heart rhythm and arrhythmia between heartbeats. Future developments of this system will take inspiration from~\cite{LVFetal2006} to incorporate non-uniform alignment between reference heartbeats for more robust phase locking.

\acknowledgments 

This work was supported by a BHF New Horizons grant (NH/14/2/31074), BHF Centre of Research Excellence Award (Edinburgh), and made use of equipment funded by EPSRC (EP/M028135/1).

\renewcommand{\bibfont}{\small}
\setstretch{0.8}
\setlength\bibitemsep{0pt}
\printbibliography[heading=bibliography]

\end{document}